\documentclass[twocolumn,showpacs,preprintnumbers,amsmath,amssymb,prb,aps]{revtex4-1}
\usepackage{epsfig}
\usepackage{epstopdf}
\usepackage{graphicx}
\usepackage{dcolumn}
\usepackage{bm}
\usepackage{textcomp}
\usepackage{array}
\usepackage{subcaption}
\usepackage{booktabs}

\begin{document}

\title{Two functionals approach in DFT for the prediction of thermoelectric properties of Fe$_{2}$ScX (X = P, As, Sb) full Heusler compounds  }
\author{Shivprasad S. Shastri}
\altaffiliation{Electronic mail: shastri1992@gmail.com}
\author{Sudhir K. Pandey}
\altaffiliation{Electronic mail: sudhir@iitmandi.ac.in}
\affiliation{School of Engineering, Indian Institute of
Technology Mandi, Kamand - 175005, India}

\date{\today}

\begin{abstract}
In the quest of new thermoelectric materials with high power factors, full-Heusler compounds having flat band are found to be promising candidates. In this direction, Fe$_{2}$ScX (X=P,As,Sb) compounds are investigated using mBJ for the band gap and SCAN to describe the electronic bands and phonon properties for thermoelectric applications. The band gaps obtained from mBJ are   0.81 eV, 0.69 eV and 0.60 eV for Fe$_{2}$ScX compounds, respectively. The phonon dispersion, phonon density of states (DOS) and partial DOS are calculated. The phonon contributions to specific heat are obtained as a function of temperature  under harmonic approximation. The electronic band structutre calculated from mBJ and SCAN functionals are qualitatively compared. The effective mass values are calculated at the band extrema from SCAN functional.  The thermoelectric parameters are calculated for both hole and electron dopings under semiclassical theory. We use simple, but reasonable method to estimate the phonon relaxation time ($\tau_{ph}$). Using the specific heat, estimated $\tau_{ph}$ and slopes (phase velocity) of acoustic branches in the linear region, lattice thermal conductivity ($\kappa_{ph}$) at 300 K is calculated for three compounds. The obtained values of $\kappa_{ph}$ with constant $\tau_{ph}$ are 18.2, 13.6 and 10.3  $Wm^{-1}K^{-1}$, respectively. Finally, the temperature dependent figure of merit $ZT$ values are calculated for optimal carrier concentrations in the doping range considered, to evaluate the materials for thermoelectric application. The $ZT$ values for n-type Fe$_{2}$ScX, in 900-1200 K, are 0.34-0.43, 0.40-0.48 and 0.45-0.52, respectively. While, the p-type Fe$_{2}$ScX have $ZT$ of 0.25-0.34, 0.20-0.28 and 0.18-0.26, respectively in the same temperature range. The $ZT$ values suggest that, Fe$_{2}$ScX compounds can be promising materials in high temperature power generation application on successful synthesis and further $\kappa_{ph}$ reduction by methods like nanostructuring.
\end{abstract}

\maketitle

\section{Introduction} 
The development of materials which can extract the heat energy and transform it into electrical energy is an important area of research.\cite{compintro,tritt} Since, devices made out of such materials can be installed at automobile heat generating parts, to convert the industrial waste heat, in home heating appliances, etc. Those materials are termed thermoelectric (TE) materials and device fabricated using them are called thermoelectric generators (TEG).\cite{yangintro} Research in the area of developing TE materials is important since they can be a very good alternative energy sources in many small scale applications. TEGs are green energy sources compared to the conventional fossil fuel based energy sources. The TE materials with dimensionless figure of merit $ZT\geq$ 1.0 are considered as  suitable for TE applications. But, the bottle neck in implementing the most of the existing TE materials into technological applications is due to their poor value of  $ZT$. Therefore, there has been continuous efforts in TE area of research to enhance the efficiency  as well as in searching for new TE materials. 

 Efficiency of a thermoelectric can be increased by maximizing it's electrical power and by reducing the heat transport by electrons and phonons. The electrical power depends on the electrical conductivity,$\sigma$, and Seebeck coefficient,$S$, of the material.\cite{yangintro} The transport of heat in a material is decided by the electronic  and phonon contributions to thermal conductivity. Thus, in order to have a thermoelectric with high figure of merit, given by the relation,
 \begin{equation}
ZT = \frac{S^{2}\sigma T}{\kappa}
\end{equation} 
the power factor, $S^{2}\sigma$ must be higher and thermal conductivity $\kappa$=$\kappa_{e}$+ should be lower. The thermal conductivity $\kappa$ in a material is a sum of electronic and lattice part of thermal conductivities denoted by $\kappa_{e}$ and $\kappa_{ph}$, respectively.  Many methods like alloying, nanostructuring are employed to reduce the lattice thermal conductivity by increasing phonon scattering without or least affecting  it's electronic structure.\cite{snydercomplex} Another method towards improving efficiency is to improve the power factor of the material. The power factor of a material is mainly decided by it's electronic structure  and such materials are being searched which could have  electronic structure that yields high power factor.\cite{mahanbest} 
 
The studies of electronic structure and thermoelectric properties carried out on many Heusler alloys suggested that, they possess a flat conduction band along $\Gamma -X$ direction and hence larger effective mass of carriers and higher Seebeck coefficient values.\cite{sharmaco2mnge,yabuuchi,flatband1,flatband2}  These works suggested that full-Heusler alloys with a semiconducting ground states could be used for thermoelectric applications. Sharma \textit{et al.} \cite{sharmafe2scx} reported three new full-Heusler alloys having flat conduction band \textit{viz.} Fe$_{2}$ScP, Fe$_{2}$ScAs and Fe$_{2}$ScSb with their electronic structure and thermoelectric properties using density functional theory (DFT) calculations. In a DFT calculation, generally band gaps are underestimated and Fe$_{2}$ScSb compound in the work of Ref.10 was predicted to be a semimetal. The transport coefficients to be calculated in first-principles calculations are highly dependent on the band gap and band structure of the compound. The electronic structure changes depending on the exchange-correlation (XC) functional used in the DFT calculations. Therefore, in the computational discovery of thermoelectric materials, selection and benchmarking of XC functional becomes an important step.

In our previous work \cite{paper1}, electronic structure of two iron based Heusler alloys Fe$_{2}$VAl and Fe$_{2}$TiSn were studied using five XC functionals. The effective mass values estimated showed the dependence of band features on XC functionals and  usefulness of mBJ in predicting accurate band gaps providing a bench mark study.  In semiconductors and insulators, transport of heat is mainly through the  phonons and thus study of phonon properties is important to understand the thermal conductivity. The phonon calculations are useful in extracting information about stability of the crystal, lattice contribution to specific heat and other thermal properties, and thermal expansion of crystals. A computational study of phonon dispersion of a material along with it's electronic structure would completely describe TE material which makes phonon calculations, desirable in a study of TE properties. Recently, in the work of Shamim \textit{et al.},\cite{shamim} the Seebeck coefficient $S$ value of Fe$_{2}$VAl was measured in the range 300-620 K. This experimental $S$ values were explained using DFT and Boltzmann transport calculations using different XC functionals. The best matching between the experimental and theoretical values was found, for using the band gap obtained from mBJ and band structure from SCAN or PBEsol. Therefore, motivated by the results of the work of Shamim \textit{et al}\cite{shamim}, we investigate the electronic structure and thermoelectric properties of the Fe$_{2}$ScX (X=P, As, Sb) compounds.

 To investigate the electronic structure and dependent properties  mBJ and SCAN  functionals are used.  Using the mBJ, band gaps of  0.81, 0.69 and 0.60 $eV$ are obtained for Fe$_{2}$ScX (X=P, As, Sb) compounds, respectively. Using SCAN, phonon dispersion , total density of states (DOS) and partial DOS are also calculated. The specific heat contribution from the lattice part of compounds are calculated.  The Debye temperatures estimated from the highest phonon frequency for three compounds are $\sim$637 K, $\sim$556 K and $\sim$498 K, respectively. Seebeck coefficient, electrical conductivity per relaxation time, power factor per relaxation time and electronic thermal conductivity per relaxation time are calculated with temperature for different values of electron and hole dopings. The effective mass values are calculated at the band extrema. The electronic band structures obtained from mBJ and SCAN are qualitatively compared. We make use of simple method in this work to estimate phonon relaxation time $\tau_{ph}$. Using calculated specific heat, $\tau_{ph}$ and slopes of acoustic branches in the linear region close to $\Gamma$, $\kappa_{ph}$ values at 300 K are calculated. Then to further evaluate the material for TE applications, we calculate temperature dependent $ZT$  for three compounds for the highest power factor giving electron and hole doping concentrations.

\begin{figure*}
\includegraphics[width=17cm, height=5cm]{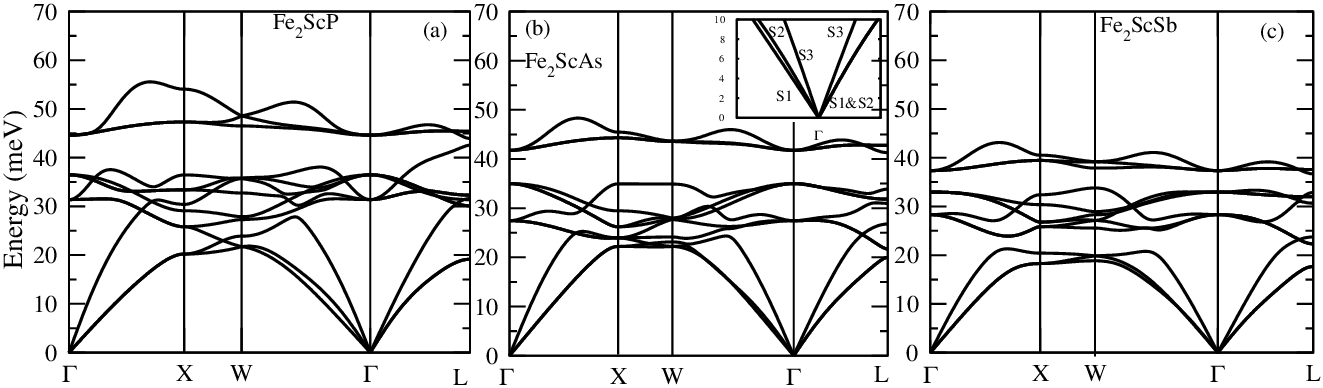} 
\caption{Phonon dispersion curves of Fe$_{2}$ScP, Fe$_{2}$ScAs and Fe$_{2}$ScSb. }
\label{fig:image2}
\end{figure*}

\section{Computational details}

In our work, full-potential linearized augmented plane wave (FP-LAPW) method based first-principles DFT program WIEN2k\cite{wien2k} is used for the calculation of ground state energy, electronic strcture and total forces on atoms. mBJ exchange potential with LDA correlation\cite{mbj} and SCAN\cite{scan} exchange-correlation functionals are chosen for the calculations. Here,  The lattice parameters are taken from the Ref. 10 and used for electronic properties calculation. Since, phonon frequency is sensitive to the lattice parameter, it is optimized for force calculations with SCAN functional . For force constants and phonon properties calculations phonopy\cite{togo} code is used. Phonon properties are calculated under finite displacement method (FDM) and supercell approach in phonopy. To capture the long-range force constant between atoms a supercell of size 2 x 2 x 2 with 128 atoms is constructed. To create the forces artificially in the system each of inequivalent atoms in the formula unit of Fe$_{2}$ScX is displaced by 0.02 Bohr in x-direction. To obtain the forces on the atoms in the created supercell, force convergence criteria of 0.1mRy/Bohr is set in the force calculator (WIEN2k). A k-mesh of size 5 x 5 x 5 is used for the force calculations in supercell. The calculations of band structure dependent transport coefficients are done using BoltzTraP program.\cite{boltztrap} The ground state total energy calculations are done on dense k-mesh of 50 x 50 x 50 size in order to facilitate transport calculations. The convergence criteria to meet the self-consistency in iterations for total energy 10$^{-4}$ Ry/cell and for charge/cell is 10$^{-2}$ electronic charges, respectively are used.

\section{Results and Discussion}
We built the crystal structure of Fe$_{2}$ScX (X=P, As, Sb) full-Heusler compounds  in the cubic L2$_{1}$ phase with sapce group $Fm-3m$\cite{sharmafe2scx}. In Ref. 10 the ground state formation energy of the static lattices have been calculated. The obtained negative values of formation energies supported the feasibility of synthesis of Fe$_{2}$ScX (X=P, As, Sb) in laboratories. Further, to study the structural stability of the compounds we have carried out phonon calculations. The phonon frequencies are sensitive to lattice parameter and exchange-correlation functional\cite{lattinflu} used in the computation (which was also observed in our previous work.\cite{paper2}) Therefore, the lattice parameters taken from Ref. 10 were optimized to get the ground state lattice parameters for SCAN functional. The equilibrium lattice parameters after fitting the B-M EOS\cite{birch} for the energy-volume curves are 5.704, 5.828, and 6.085 {\AA} for Fe$_{2}$ScP, Fe$_{2}$ScAs and Fe$_{2}$ScSb, respectively. Using these lattice parameters from SCAN phonon dispersion, density of states (DOS) and electronic dispersions are studied.

\begin{figure*}
\includegraphics[width=16cm, height=6cm]{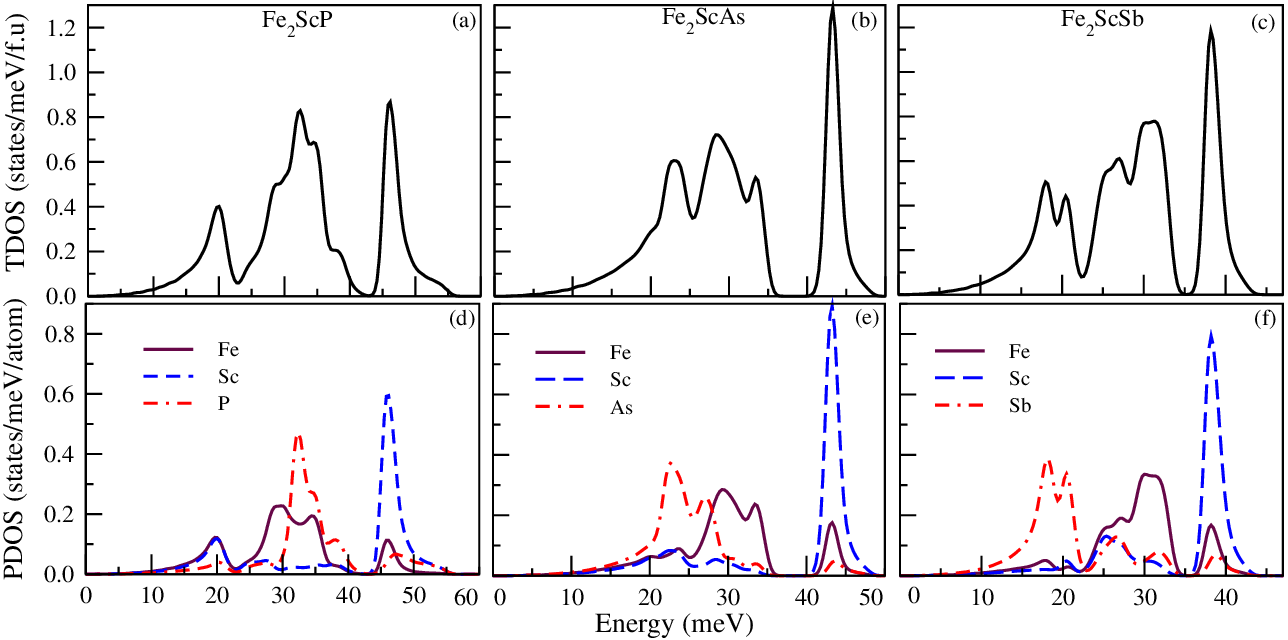} 
\caption{Phonon TDOS and PDOS of Fe$_{2}$ScP, Fe$_{2}$ScAs and Fe$_{2}$ScSb.}
\end{figure*}

\subsection{\label{sec:level2}Phonon properties}
The phonon dispersions for Fe$_{2}$ScX (X=P, As, Sb) compounds are presented in Fig. 1(a) to (c). The calculated phonon spectrum does not show any negative frequency (or energy) which suggests that crystal is stable for any small atomic displacements about the mean position. Dynamical stability of the compounds found out with phonon calculations and formation energy values from Ref. 10 further supports the possibility of preparation of samples in laboratories.

In the figure, one can observe that maximum phonon energy is decreasing from Fe$_{2}$ScP to Fe$_{2}$ScSb. The highest energy phonons are of $\sim$55, $\sim$48 and $\sim$43 meV for Fe$_{2}$ScP, Fe$_{2}$ScAs and Fe$_{2}$ScSb, respectively. The observed decreasing trend in phonon frequency can be mainly due to the following reasons. The radius of atoms X (P, As, Sb) in the unit cell of Fe$_{2}$ScX is increasing from P to Sb making distance between atoms and hence, the lattice constant to increase from Fe$_{2}$ScP to Fe$_{2}$ScSb.  Also, the X atoms in the unit cell are heavier in mass on going from P to Sb. Since, the mass of atoms and distance between them are increased the frequency of vibrations of atoms is expected to reduce from first to last compound in the Fe$_{2}$ScX family.    The phonon properties and electronic structure for Fe$_{2}$ScAs are discussed in our earlier work \cite{dae} but, to study the family of Fe$_{2}$ScX with X = P, As, Sb atoms for thermoelectric applications, we included Fe$_{2}$ScAs here, for the purpose of completeness in the discussion. 

The phonon dispersion for Fe$_{2}$ScP in Fig. 1 (a) shows three acoustic branches, out of which one branch is degenerate with optical branch at k-points nearer to $X$-point in the $\Gamma$-$X$ direction. Three optical branches starting at 45 meV are well seperated from the rest of the branches along the k-path directions shown in the plot except near $L$-point, around 43 meV, where the seperation is very much reduced ($\sim$0.14 meV). This seperation is also noted in the dispersions of  Fe$_{2}$ScAs and Fe$_{2}$ScSb. In Fe$_{2}$ScAs above 40 meV and in Fe$_{2}$ScSb above 35 meV optical branches with clear seperation are situated. In case of Fe$_{2}$ScSb, optical and acoustic branches are touching only near $L$-point at $\sim$22 meV unlike in other two compounds. From phonon dispersion we can also observe that the maximum energy for acoustic phonons in a given compound is reducing from Fe$_{2}$ScP to Fe$_{2}$ScSb. 

Fig. 2(a)-(c) and (d)-(f) show the calculated phonon total density of states (TDOS) and partial DOS for Fe$_{2}$ScX (X=P, As, Sb), respectively. From the atom specific phonon DOS of Fe$_{2}$ScP (Fig. 2(d)), we can see that higher energy optical phonon branches ($\sim$45-55 meV) are contributed mainly from Sc atom. The considerable number of states for phonons due to the vibrations of lighter mass P atom is in the energy range $\sim$30-40 meV. The lower energy acoustic phonons in the range $\sim$10-22 meV are mainly due to the Sc and Fe atoms in the unit cell. The contributions to different branches in the phonon dispersion from the atoms in the unit cell of Fe$_{2}$ScAs can be understood from phonon PDOS in Fig. 2(e). The higher energy states of optical phonons in $\sim$40-50 meV are mainly contributed from lighter Sc atom in the formula unit. The major number of vibrational states lying in the intermediate energy region $\sim$25-30 meV   can be attributed  to Fe and As atoms.  For Fe$_{2}$ScSb, TDOS and PDOS of phonon states are shown in Fig. 2(c) and (f), respectively. The heavier Sb atom is contributing considerably to the phonon states in $\sim$10-22 meV region. Like in the case of Fe$_{2}$ScAs,  there is a small region in the neighborhood of 35 meV with no phonon states. The higher energy states of optical phonon states ($\sim$36-43 meV) are primarily due to the Sc atoms (Fig. 2(f)). Fe atom contribution is mainly to phonons of energy $\sim$22-34 meV. The vibrational contribution of atoms X (X=P, As, Sb) in the unit cell of Fe$_{2}$ScX is shifting towards to the phonons of lower energy on crossing to Fe$_{2}$ScSb from Fe$_{2}$ScP. The increase in the atomic mass of X elements as well as increase in lattice constant mainly being the reason for this observed shift. But, in all the three cases the peaks in the higher energy range in DOS plots corresponding to three optical branches are substantially due to vibrations of the Sc atom. 
\begin{figure}
\vspace{1.8cm}
\includegraphics[width=8cm, height=5cm]{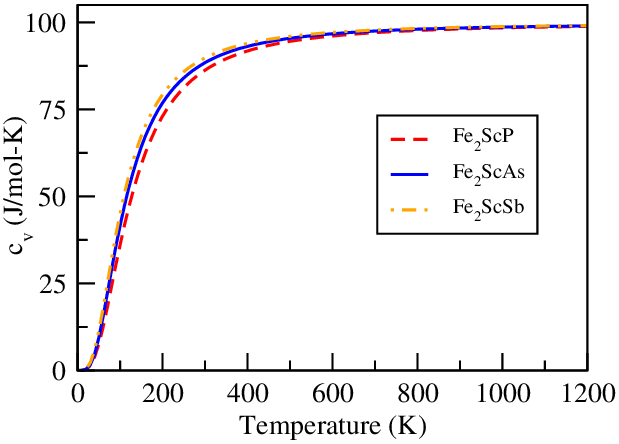} 
\caption{Constant volume specific heat $c_{v}$  curves of Fe$_{2}$ScP, Fe$_{2}$ScAs and Fe$_{2}$ScSb.}
\end{figure}

\begin{table}[b]
\caption{\label{tab:table1}%
The indirect band gaps in $eV$ obtained from mBJ, SCAN and PBEsol functionals.\cite{sharmafe2scx}
}
\begin{ruledtabular}
\begin{tabular}{lccc}
\textrm{Functional}&
\textrm{Fe$_{2}$ScP}&
\textrm{Fe$_{2}$ScAs}&
\textrm{Fe$_{2}$ScSb}\\
\colrule
mBJ & 0.81 & 0.69 & 0.60\\
SCAN & 0.31  & 0.14 & 0.06  \\
PBEsol & 0.30 & 0.09 & -\\
\end{tabular}
\end{ruledtabular}
\end{table}

The lattice contributions to the constant volume specific heat $c_{v}$ of Fe$_{2}$ScX compounds calculated under harmonic approximation are shown in Fig. 3. At higher temperatures (above 800 K) $c_{v}$ curves are approaching the classical Dulong and Petit limit  of $\sim$100 J/mol-K for Fe$_{2}$ScX compounds. At any temperature, one can see that  $c_{v}$ values for Fe$_{2}$ScSb $>$ Fe$_{2}$ScAs $>$ Fe$_{2}$ScP compound. The lattice part of thermal conductivity is directly proportional to $c_{v}$ and hence calculation of $c_{v}$ is essential for an estimation of $\kappa_{ph}$ of a TE material.  From the nature of the curves, one can estimate that at higher temperatures the differences in $\kappa_{ph}$ of Fe$_{2}$ScX compounds should be mainly due to group velocity and phonon relaxation time rather than due to  $c_{v}$ which are closer to each other in this case. Considering the definition of Debye frequency as a measure of the maximum phonon frequency and Debye temperature, $\Theta_{D}$, to be the temperature above which all phonon modes begin to be excited\cite{ashcroft}, the calculated values of $\Theta_{D}$ for Fe$_{2}$ScX (X=P, As, Sb) are $\sim$637 K, $\sim$556 K and $\sim$498 K, respectively. The maximum phonon energy is decreasing from Fe$_{2}$ScP to Fe$_{2}$ScSb as mentioned earlier which is the reason for observed trend of $\Theta_{D}$.

\subsection{\label{sec:level2}Electronic structure}
The band gaps of semiconductors and insulators in KS-DFT calcualtions are generally underestimated. But, in a theoretical study or in prediction of new compounds with a band gap for thermoelectric applications, accurate calculation of band gap is essential since the number of carriers thermally excited varies exponentially with band gap. Also, in a DFT calculation, the curvature of bands at the band extrema and band features of electronic dispersion depends on the approximation used for the exchange-correlation part.\cite{paper1} The transport coefficients calculated, depend on the effective mass and group velocity, which are derived from the band structure. Therefore, proper selection of XC functional for any DFT study of TE material is crucial. The question of which functional to choose was addressed in the work of Shamim \textit{et al.}\cite{shamim} as mentioned in the introduction which proposed that the combination of mBJ and SCAN (or PBEsol) functional would better explain the Seebeck coefficient. Thus, to calculate electronic band structure and dependent transport properties, two functionals \textit{viz.} SCAN and mBJ are used. 
\begin{figure*}
\includegraphics[width=17cm, height=8cm]{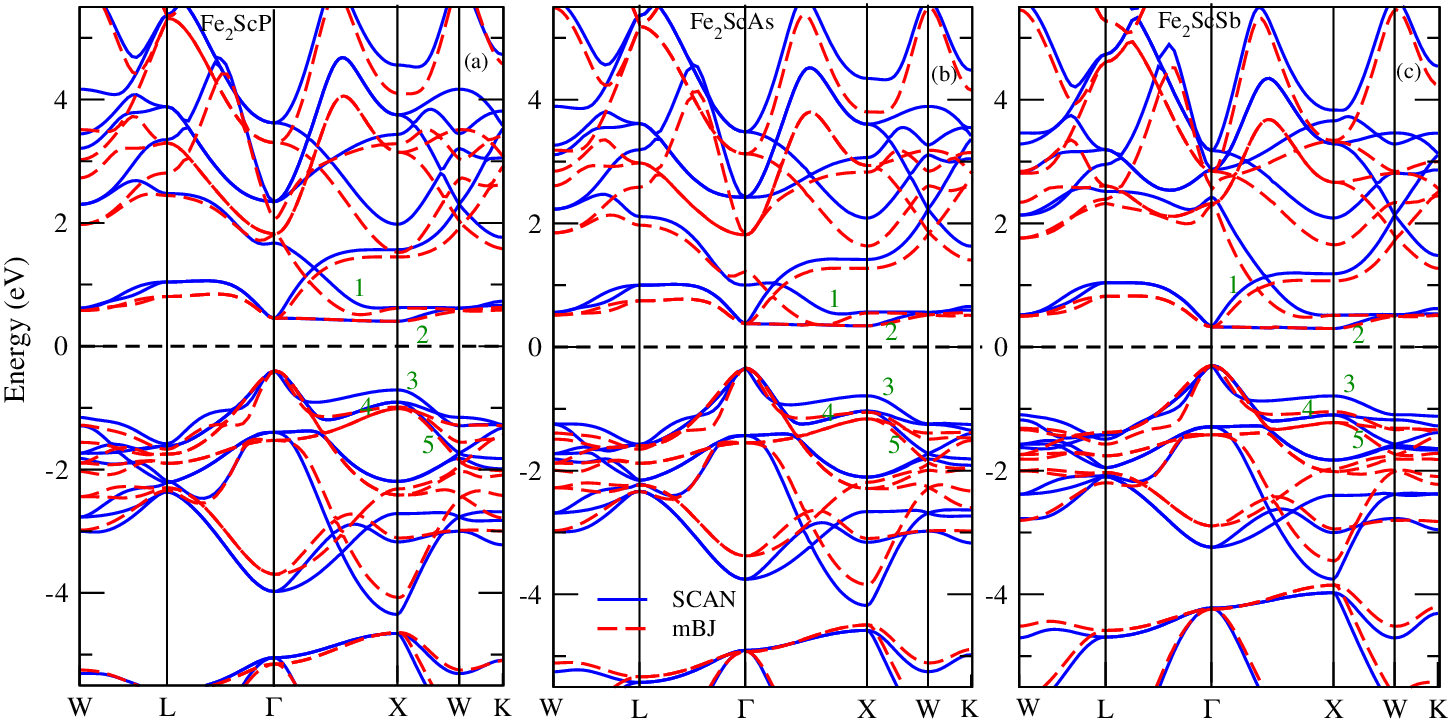} 
\caption{Electronic band structures of Fe$_{2}$ScP, Fe$_{2}$ScAs and Fe$_{2}$ScSb. The blue (solid) lines represent bands from SCAN and red (dashed) lines are from mBJ.}

\end{figure*}
The electronic bandstructure of Fe$_{2}$ScX (X=P, As, Sb) compounds are shown in Fig. 4(a)-(c), respectively. In the plots, the zero energy corresponds to the Fermi level, $E_{F}$, and is set to the middle of the gap.  For each compound the indirect band gaps obtained from both functionals  are listed in Table I. The indirect band gaps of Fe$_{2}$ScP and Fe$_{2}$ScAs obtained in the work of Sharma \textit{et al.}\cite{sharmafe2scx} are also included for comparison. Fe$_{2}$ScSb was predicted as semimetallic while former two compounds as semiconducting from PBEsol calculations in Ref. 10.  As can be seen in Table I, the band gap values predicted by SCAN and PBEsol are closer to each other but, much lesser than that from  mBJ. The band gaps of Fe$_{2}$ScP, Fe$_{2}$ScAs and Fe$_{2}$ScSb are 0.81, 0.69 and 0.60 $eV$, respectively as obtained from mBJ calculations. The mBJ  is constructed to give accurate band gaps and the literatures show the good agreement between the experimental and mBJ calculated band gaps.\cite{kim,paper1} Thus, band gaps predicted in this work could be useful numbers to experimentalists for comparison in case samples are prepared. 

The  band gap is reducing from  Fe$_{2}$ScP to Fe$_{2}$ScSb by a value of approximately 0.1 eV in case of mBJ. Therefore, at a given temperature the number of electrons excited into the conduction band are relatively lower in Fe$_{2}$ScP compared to other two compounds with the relatively least number being in Fe$_{2}$ScSb.  A band gap of mBJ ground state band structure has been introduced into the band structure of SCAN  in such a way that conduction band minimum (CBM) at $X$-point and valence band maximum (VBM) at $\Gamma$-point from both functionals coincide. This representation of band structure from two functionals depicts (Fig. 4) the  differences in the features of dispersion curves at different points along the k-path. The bands numbered 1 to 5 in Fig. 4 are denoted with symbols B1 to B5 in the description.  The X atoms in the compound does not found to significantly affect the electronic states in the $\sim$-4 to -0.3 eV energy range of valence band region. Similarly, bands B1 and B2 in CB region  are also slightly altered.  This is discernible in the plots since the features of bands qualitatively does not show much difference in energy range mentioned.

 In the plots the valence band maximum (VBM) is at $\Gamma$-point and CBM is at $X$-point giving indirect band gaps. In all three compounds, the VBM is triply degenerate at $\Gamma$-point. Fe$_{2}$ScX compounds possess a band B2 which is almost flat along the $\Gamma$-$X$ direction and is dispersive in the direction $\Gamma$-$L$-$W$.  The bands B1 and B2 are degenerate at $\Gamma$-point and continues to show degeneracy  till $L$-point. The  bands B1 and B2 start at the same energy at $W$ but along the $W$-$\Gamma$ and $\Gamma$-$X$ directions, the bands of SCAN are elevated in energy relative to the mBJ bands. The three valence bands of SCAN starting from $\Gamma$ and along the directions $L$-$W$ and $X$-$W$-$K$  are higher in energy relative to  mBJ bands.  In other words, in the VB region nearer to $E_{F}$, SCAN bands are narrower compared to mBJ bands (B3-B5) which implies density of states should be higher near $E_{F}$ in electronic structure calculated from SCAN . The CBM at $\Gamma$-point is doubly degenerate, but the bands 1 and 2 obtained from mBJ are relatively more narrower compared to the bands of SCAN. This implies the density of states in the CB region near $E_{F}$ is more in mBJ approximated electronic ground state.

The relatively small number of electrons excited into the conduction band decide the transport behavior in a semiconductor.\cite{ashcroft} Since, the excited carriers are found almost in the neighborhood of the CB minima or VB maxima, for the three compounds the band extrema are approximated with a parabola to calculate the effective mass. Under the parabolic approximation, effective mass is inversly proportional to the curvature of the parabolic band. Thus, the effective mass values of carriers in a band indicates the curvature of the band. The Seebeck coefficient (S) is directly proportional to the effective mass of charge carriers in the material according to the equation\cite{snydercomplex}, 
\begin{equation}
S = (8\pi^{2}k^{2}_{B}/3eh^{2})m^{*}T(\pi/3n)^{2/3}.
\end{equation} 
In the above relation, $h$ is Planck's constant, $k_{B}$ is Botlzmann constant, $e$ is electronic charge  and $n$ is carrier concentration, respectively. 

 The direct band gap from mBJ in these materials are 0.85, 0.72 and 0.62 eV, respectively (Fig. 4).  These values of energy gap are much close to the indirect band gaps of mBJ listed in Table I and hence there is finite probability that electron transition takes place between these points too. In order to quantify the curvature of bands from SCAN, the parabola were fitted at band extrema according to the free electron energy equation $E=\frac{\hbar^{2}k^{2}}{2m_{e}m^{*}}$, where $m_{e}$ is rest mass of electron and $m^{*}$ effective mass in terms of mass of electron.  The bands numbered 1 to 5 (labeled B1 to B5 in Table II) in Fig. 4 which significantly participate in the transport are fitted with parabola in the vicinity of CB minima at $\Gamma$ and $X$-points and VB maximum at $\Gamma$-point  for SCAN functional. The obtained values of effective mass along different directions for Fe$_{2}$ScX (X=P, As, Sb) for SCAN bands are listed in Table II.  The notation, for instance, $\Gamma$ - $\Gamma L$ denotes the effective mass calculated at $\Gamma$-point along $\Gamma$ to $L$ direction.

\begin{figure*}
\vspace{1.5cm}
\includegraphics[width=17cm, height=11cm]{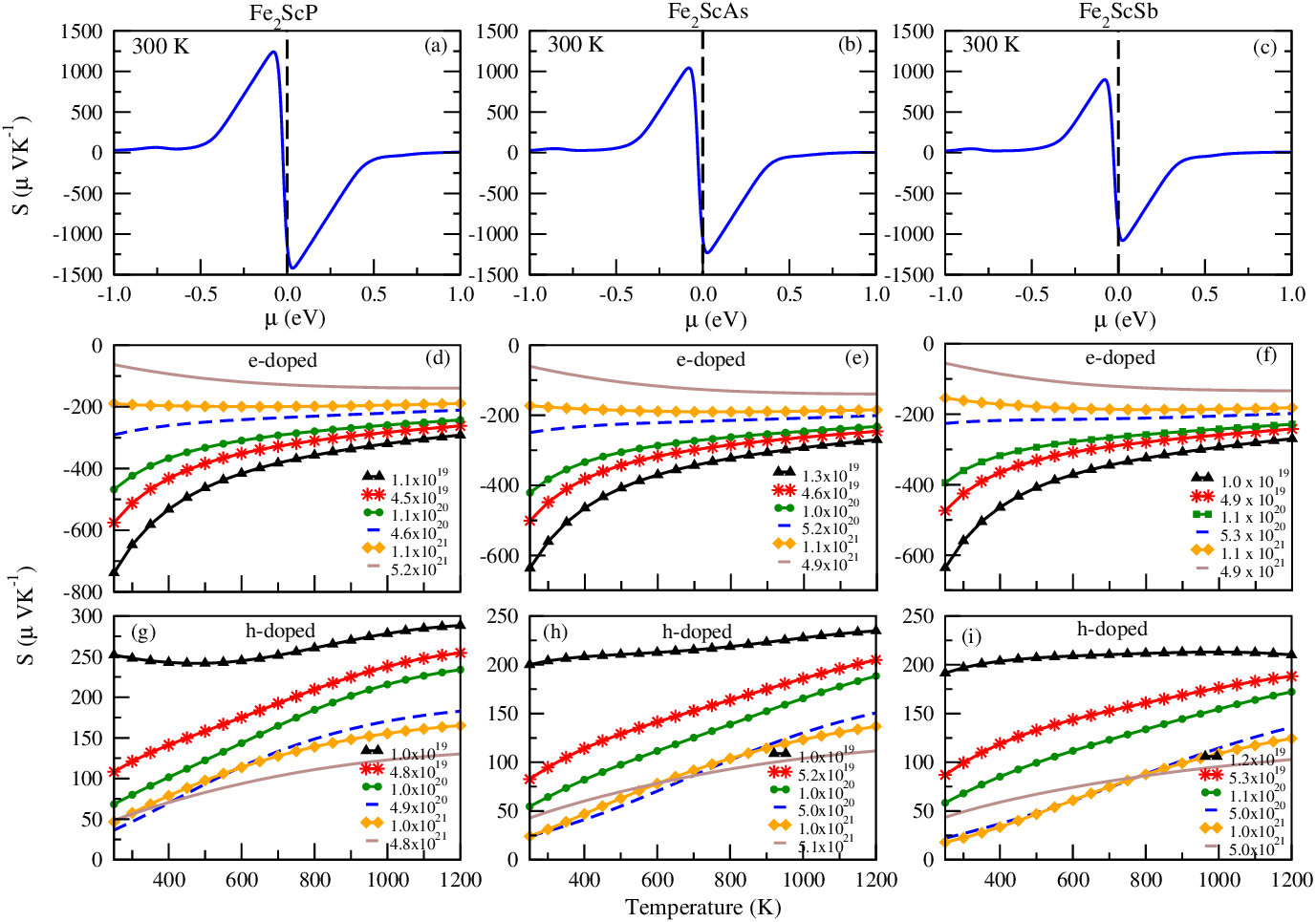} 

\caption{Seebeck Coeffcient (S) vs. chemical potential ($\mu$) plots (Fig. 5(a)-(c)) and S as a function of temperature (T) for electron doping (Fig. 5(d)-(f))and hole doping (Fig. 5(g)-(i)). Each of three vertical panel in the figure shows the plots for Fe$_{2}$ScP, Fe$_{2}$ScAs and Fe$_{2}$ScSb, respectively.}
\end{figure*}

\begin{table*}
\caption{Effective mass($m^{*}$) of charge carriers calculated for Fe$_{2}$ScX (X=P, As, Sb  compound at $\Gamma$ and X points respectively, in various bands from SCAN.}
\resizebox{0.8\textwidth}{!}{%
\begin{tabular}{@{\extracolsep{\fill}} c c c c c c c c c c c c c c c}
 \hline\hline

 & \multicolumn{4}{c}{Fe$_{2}$ScP} &  & \multicolumn{4}{c}{Fe$_{2}$ScAs} &  & \multicolumn{4}{c}{Fe$_{2}$ScSb}\\
 \cline{2-5} \cline{7-10} \cline{12-15} 
 &$\Gamma$-$\Gamma$L &$\Gamma$-$\Gamma$X  &X-X$\Gamma$ &X-XW   &  &$\Gamma$-$\Gamma$L &$\Gamma$-$\Gamma$X  &X-X$\Gamma$ &X-XW  & &$\Gamma$-$\Gamma$L &$\Gamma$-$\Gamma$X  &X-X$\Gamma$ &X-XW\\
\hline
B1   &0.26 &0.13   &       &     &  &0.19   &0.10    &      &     &  &0.16 &0.09   &       &  \\
B2   &0.26 &20.56     &39.91  &1.36 &  &0.19   &46.79 &36.67 &1.46 & &0.16  &50.25 &37.69 &1.25\\
B3   &0.26 &0.43 &       &     &  &0.21   &0.41 &       &    &  &0.22 &0.45  &       & \\
B4   &0.26 &0.42 &       &     &  &0.19   &0.41 &       &    &  &0.16 &0.45  &       & \\ 
B5   &0.22 &0.14 &       &     &  &0.19   &0.10 &      &     &  &0.16 &0.09  &       & \\  
\hline
\end{tabular}}
\end{table*} 
    The effective mass values in Table II clearly highlight the large $m^{*}$ for band B2 in all three compounds. Along $X-W$ direction $m^{*}$ are 39.91, 36.67 and 37.69  for three compounds, respectively. For the same band along $\Gamma$-$X$ direction carriers have $m^{*}$ greater than along $X-W$ direction.  In Fe$_{2}$ScP, effective mass $m^{*}$ of the bands B1-B4 is 0.26 and B5 is 0.22, respectivley along $\Gamma$-$L$ directions. Along the $\Gamma$-$X$ direction $m^{*}$ of bands B1, B5, and B3 and B4 are 0.13, 0.14, 0.43 and 0.42, respectively.  As can be seen from the table, the band B2 has very large value of effective mass ($m^{*}$) along $\Gamma-X$ and $X-\Gamma$ directions and can be called heavy band in  Fe$_{2}$ScX. From Fig. 4 one can observe that at CBM and VBM at the $\Gamma$-point, the curvature of mBJ bands are lesser compared to that of SCAN bands.  This implies that effective mass if estimated from mBJ should be higher for these compounds. In the work of  Kim \textit{et al.} \cite{kim} for III-V semiconductors effective mass values were estimated using mBJ and found that mBJ effective masses are overestimated with respect to experimental data by $\sim$30-50 \%. Thus, one may predict that effective mass calculated for SCAN should be nearer to experimental value since these band features also explained the Seebeck coefficient of Fe$_{2}$VAl in Ref. 12.  For a given compound, along the $\Gamma - L$ direction the bands B1 to B5 have nearly same curvature. While, along $\Gamma - L$ direction, B3 and B4 bands are of similar curvature which is justified by the $m^{*}$ values in the table.

\subsection{\label{sec:level2}Thermoelectric properties}
The thermoelectric parameters Seebeck coefficient $S$, electrical conductivity per relaxation time $\sigma/{\tau}$ and power factor per  relaxation time $S^{2}\sigma/{\tau}$ (PF), are calculated for Fe$_{2}$ScX compounds under semiclassical transport theory. In order to calculate TE properties, the band structure from SCAN functional and band gap obtained from the mBJ have been used. In Fig. 5(a)-(c), for three compounds the variation $S$ with the shifts in chemical potential ($\mu$) are presented at 300 K. The dashed line normal to $\mu$ axis indicates chemical potental for the intrinsic compounds at 0 K. The positive and negative values of $\mu$ in plots correspond to electron and hole doping, respectively. At the temperature of 300 K,  $S$ vs. $\mu$ curve cuts the $\mu$=0 line, for Fe$_{2}$ScP, at $\sim$-1134 $\mu$VK$^{-1}$ and for Fe$_{2}$ScAs and Fe$_{2}$ScSb at $\sim$-1068 and $\sim$-963 $\mu$VK$^{-1}$, respectively. This indicates intrinsic Fe$_{2}$ScX at 300 K will show negative $S$ values which is attributed mainly due to presence of flat conduction band in these compounds.  The values of extrema in the $S$ curves are reducing from Fe$_{2}$ScP to Fe$_{2}$ScSb which implies in Fe$_{2}$ScP maximum  $S$ can be obtained either by electron or hole doping. 

 The maximum value of $S$ (at 300 K) for the three compounds are $\sim$-1421, $\sim$-1230 and $\sim$-1080 $\mu$VK$^{-1}$ in the order, for shift in $\mu$ of $\sim$34, $\sim$23 and $\sim$24 meV ,respectively. While, the maximum positive $S$ values that can be attained by hole doping is relatively lesser compared to electron doping and also requires more doping (Fig. 5(a)-(c), large shift in $\mu$ towards negative side) to get high $S$ value.   At a given temperature, the $S$ is inversly proportional to carrier concentration ($n$) and directly proportional to $m^{*}$.  The band gap of Fe$_{2}$ScP is 0.81 eV and is the highest in the family of Fe$_{2}$ScX (Table I) making $n$ to be smaller compared to other compounds at a given temperature. This explains the reason for the observed higher $S$ in Fig. 5(a) for Fe$_{2}$ScP compared to other two compounds. We compare the $S$ vs. $\mu$ curves with the work of Sharma \textit{et al.} \cite{sharmafe2scx} wherein PBEsol functional was used. We observe that the extrema of $S$ values calculated as a function of $\mu$ at 300 K are very much lower compared to that of our work. The maximum value of $S$ in Ref. 10 for undoped compounds (at $\mu$=0) was -770, -386, -192 $\mu$VK$^{-1}$ for Fe$_{2}$ScP, Fe$_{2}$ScAs and Fe$_{2}$ScSb, respectively. This can be mainly due to the lower band gap predicted by PBEsol (Table I) for Fe$_{2}$ScX compounds.

In order to find out the proper carrier concentration that would yield $S$ and $\sigma/{\tau}$ values which maximize the PF, we have  doped Fe$_{2}$ScX compounds  with electrons and holes in the range $\sim$1x10$^{19}$ to $\sim$5x10$^{21}$ cm$^{-3}$. For the three compounds $S$, $\sigma/{\tau}$ and PF values are calculated for various dopings in the temperature range 250-1200 K.  Seebeck coefficient values for electron doped compounds are shown in Fig.5 (d)-(f) . For Fe$_{2}$ScX, values of $S$ are decreasing continuously with the increase in the temperature for a given electron doping from $\sim$1x10$^{19}$ cm$^{-3}$ to $\sim$1x10$^{20}$ cm$^{-3}$. For $\sim$5x10$^{20}$ cm$^{-3}$ of doping, the range  of $S$ (max. $S$-min.$S$) for temperature range under study is reduced. For  $\sim$1x10$^{21}$ cm$^{-3}$  of doping the $S$ is not varying substantially with temperature. But, slow increasing nature of $S$ is observed for $\sim$5x10$^{21}$ cm$^{-3}$. This is the general trend of $S$ observed in case of electron doping. Also, with increase in the electron doping $S$ curves are shifting towards lower values since $S$ varies inversly with carrier concentration as given by Eq. (2). The maximum values of $S$ are $\sim$-743, $\sim$-645, $\sim$-637 $\mu VK^{-1}$  for $\sim$1x10$^{19}$ cm$^{-3}$ electron doping for Fe$_{2}$ScX compounds, respectively at 250 K. At any temperature lower doping of $\sim$1x10$^{19}$ cm$^{-3}$ has higher $S$ value. In Ref. 10, nature of the $S$ curves for $\sim$1x10$^{19}$-$\sim$1x10$^{21}$ cm$^{-3}$ of doping are different which can be due the lower band gaps considered compared to present work.  The maximum $S$ values at $\sim$1x10$^{19}$ cm$^{-3}$ doping are $\sim$-500, $\sim$-400 and $\sim$25 $\mu VK^{-1}$ for Fe$_{2}$ScX, respectively.

 The nature of $S$ curves for different hole dopings are shown in Fig. 5(g)-(i) for three compounds. The maximum $S$ is obtained for doping of $\sim$1x10$^{19}$ cm$^{-3}$ in the temperature range considered and the increment in $S$ with the increase in the temperature is less compared to higher dopings. For  $\sim$5x10$^{19}$-$\sim$5x10$^{21}$ cm$^{-3}$ range, $S$ is increasing with nature of  $S$ almost linear in 250-1200 K. But, Seebeck coefficients for dopings of $\sim$1x10$^{21}$ cm$^{-3}$ and above are crossing lower doping curves at higher temperatures. In Ref. 10, $S$ curves for hole doping does not show linear increase behavior as in this work, instead there are peaks in the $S$ curves. The reason for the observed differences may be due to the large band gap considered in this work. Comparing the $S$, for hole and electron doped case, one can observe that from Fig. 5, for a given doping electron doping gives larger $S$ for Fe$_{2}$ScX. 

The Figure 6 shows the $\sigma/{\tau}$ and  PF plots for different carrier concentrations as a function of temperature. The electron and hole doping values considered for both  $\sigma/{\tau}$ and PF are shown  only in the respective PF plots, represented in same colors (or symbols). The highlighting feature of $\sigma/{\tau}$ plots (Fig. 6(a)-(f)) is that, for any doping values  $\sigma/{\tau}$ curves are nearly linearly increasing from 250 to 1200 K for all compounds. For doping of $\sim$1x10$^{19}$-$\sim$1x10$^{20}$ cm$^{-3}$, $\sigma/{\tau}$ curves are close to each other for both electron and hole doping. For higher dopings $\sigma/{\tau}$ curves are shifting towards higher values. The highest $\sigma/{\tau}$  is obtained  for doping concentration of $\sim$5x10$^{21}$ cm$^{-3}$ of both electron and hole type. From $\sim$5x10$^{20}$ to $\sim$5x10$^{21}$ cm$^{-3}$, the obtained value of $\sigma/{\tau}$, for a given doping concentration is higher for hole type doping compared to electron type. This can be understood as because of the lesser effective mass of holes in valence bands (B3-B5)(Table II) compared to that of electrons with high effective mass in band B2. Since, the charge carrier with higher effective mass has lower mobility and electrical conductivity in semiconductors is related to mobility as, $\sigma$=$ne\mu_{c}$, where $\mu_{c}$ is mobility of charge carriers, higher electrical conductivity of holes is expected in these compounds. 

\begin{figure*}
\includegraphics[width=17cm, height=13cm]{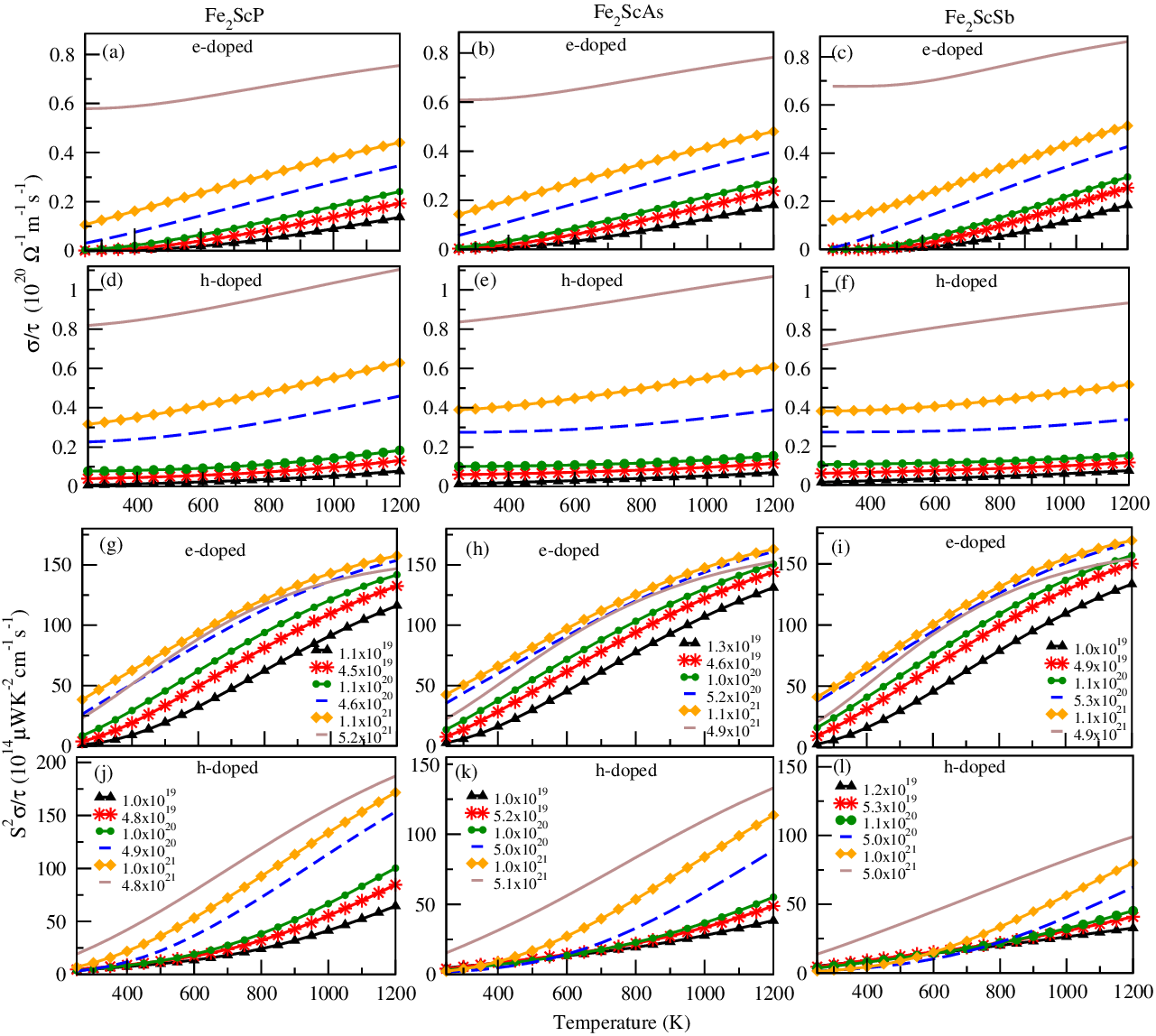} 
\caption{Electrical conductivity per relaxation time (${\sigma}/{\tau}$ )  and power factor per relaxation time ($S^{2}\sigma/{\tau}$) as a function of temperature (T) for electron doping (Fig. 6(a)-(c) and (g)-(i)) and hole doping (Fig. 6(d)-(f) and (j)-(l)) for Fe$_{2}$ScP, Fe$_{2}$ScAs and Fe$_{2}$ScSb, respectively.}
\end{figure*}

 The $S^{2}\sigma/{\tau}$ (PF) plots for Fe$_{2}$ScX are presented in Fig. 6 (g)-(l) for electron and hole doping in $10^{14}\mu WK^{-2}cm^{-1}s^{-1}$ (= 1 PF unit, PFU). For electron doped cases, the PF curves are shifted to higher values with the increased doping concentration except for doping of $\sim$5x10$^{21}$ cm$^{-3}$. For a given doping, the PF  is increasing with temperature  from 250 to 1200 K. For temperatures of practical applications, for instance, in automobiles where temperature of heat source, to install TEG is considered as 800 K \cite{gaurav}, PF values at these temperatures are, $\sim$121, $\sim$125 and $\sim$131 PFU, for doping of $\sim$1x10$^{21}$ cm$^{-3}$ for Fe$_{2}$ScP, Fe$_{2}$ScAs, and Fe$_{2}$ScSb, respectively. Thus, for Fe$_{2}$ScX compounds, to get a high PF from electron doping, $\sim$1x10$^{21}$ cm$^{-3}$ can be considered as optimal doping concentration in the doping range considered.

 For hole doped case also an increasing trend in PF with temperature is observed in Fig. 6 (j)-(l).  For doping range $\sim$1x10$^{19}$-$\sim$1x10$^{20}$ cm$^{-3}$, the PF values are close to each other. At any temperature,  for the hole doping of $\sim$5x10$^{21}$ cm$^{-3}$ highest PF is obtained for these compounds. At 800 K, the values of PF for $\sim$5x10$^{21}$ cm$^{-3}$ dopings are, $\sim$118, $\sim$83 and $\sim$63 PFU for Fe$_{2}$ScP, Fe$_{2}$ScAs, and Fe$_{2}$ScSb, respectively. From Fig. 6 (j)-(l) we can observe that PF values for a given doping is decreasing from Fe$_{2}$ScP to Fe$_{2}$ScSb. The power factor curves in the work of Ref. 10 shows gradual incresing behavior for the carrier concentration of $\sim$1x10$^{17}$ to $\sim$1x10$^{19}$ cm$^{-3}$. But, for doping of $\sim$1x10$^{20}$ to $\sim$1x10$^{21}$ cm$^{-3}$, there are peaks in the PF curves. The maximum PF is obtained for a doping of $\sim$5x10$^{21}$ cm$^{-3}$ in all three compounds. The maximum values of PF obtained for p-type doping is $\sim$90-24 PFU and for n-type doping is $\sim$85-54 PFU from Fe$_{2}$ScP to Fe$_{2}$ScSb compound in Ref. 10. The values of maximum PFs obtained by using PBEsol functional in the work of Ref. 10 are lesser compared to the present work. The band gaps obtained by PBEsol are lesser compared to mBJ as can be seen in Table I, may be the reason for the observed differences in PF values calculated in two works.  From the nature of the PF curves (Fig. 6 (g)-(l)), which are showing  increasing behaviour with temperature, we can propose these materials for high temperature thermoelectric applications. 

\begin{figure*}
\vspace{1.5cm}
\includegraphics[width=17cm, height=7cm]{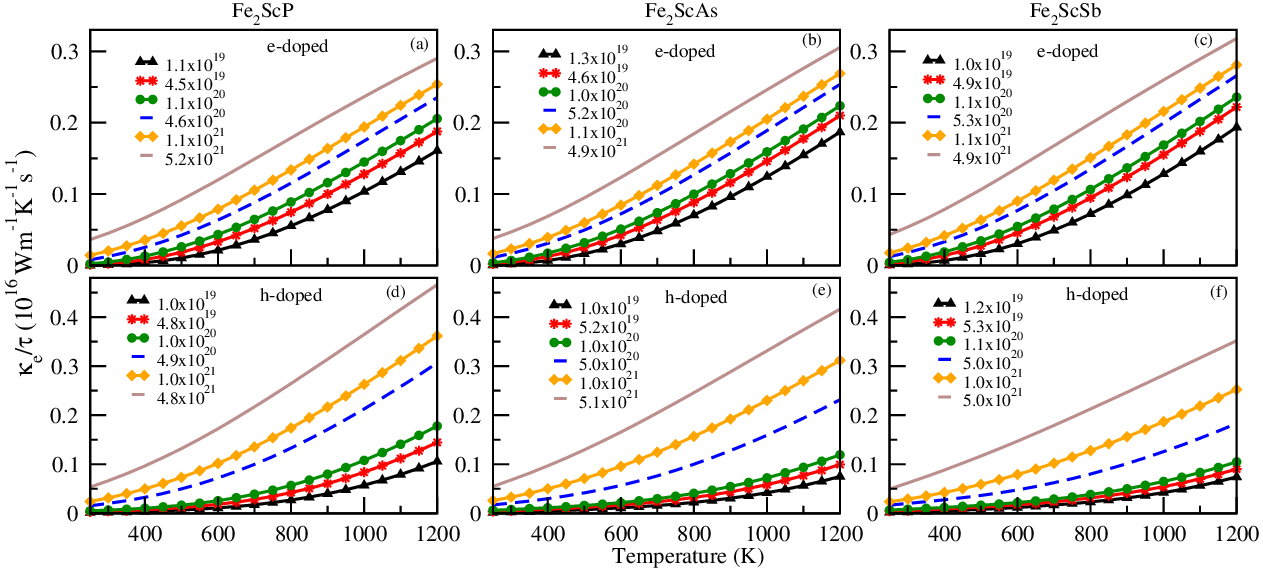} 

\caption{Electronic thermal conductivity per relaxation time (${\kappa_{e}/{\tau}}$) as a function of temperature (T) for electron doping (Fig. 7(a)-(c)) and hole doping (Fig. 7 (d)-(f)) for Fe$_{2}$ScP, Fe$_{2}$ScAs and Fe$_{2}$ScSb, respectively.  }

\end{figure*}

Electronic thermal conductivity per relaxation time (${\kappa_{e}/{\tau}}$) of Fe$_{2}$ScX compounds for electron and hole dopings considered in this study are shown in Fig. 7 (a)-(c) and (d)-(f), respectively. For three compounds, ${\kappa_{e}/{\tau}}$ curves are showing  increasing trend with temperature. For optimal doping cases giving the highest PFs $viz.$ $\sim$1x10$^{21}$ cm$^{-3}$ of electron doping,  ${\kappa_{e}/{\tau}}$ is increasing from first to last compound while, for hole doping of $\sim$5x10$^{21}$ cm$^{-3}$ reverse trend is observed. At 800 K, ${\kappa_{e}/{\tau}}$ values for electron doping of $\sim$1x10$^{21}$ cm$^{-3}$ which yields the highest PF are 0.13, 0.14 and 0.15 $10^{16} Wm^{-1}K^{-1}s^{-1}$ for Fe$_{2}$ScX compounds, respectively.

Calculation of $\kappa$ of TE material is important to fully evaulate it for TE application. But, calculation of $\kappa_{ph}$ is a difficult and computationally demanding task\cite{phono3py}. Therefore, any simple  method predicting fairly good value of $\kappa_{ph}$ is of much help. Here for a quantitative estimation of $\kappa_{ph}$ in  Fe$_{2}$ScX compounds we resort to a simple method which turns out to provide reasonably good estimation of $\kappa_{ph}$. The relation for $\kappa_{ph}$ is given by\cite{ashcroft},
\begin{equation}
\kappa_{ph} = \frac{1}{3}c_{v}c^{2}\tau_{ph},
\end{equation}
 where $c_{v}$ is constant volume specific heat per unit volume, $c$ is phonon phase velocity in linear dispersion and $\tau_{ph}$ is phonon relaxation time. 
 
 Now, to calculate $\kappa_{ph}$ from Eq. 3, the values of $c_{v}$, $c$ and $\tau_{ph}$ are needed. The specific heat, $c_{v}$ is obtained as function of temperature from the phonon calculations. To calculate $c$, three acoustic branches are approximated with linear dispersion, by fitting with straight line in the neighborhood of $\Gamma$ ($\bf{k}=0$) of the calculated phonon dispersions. Then $c$ is calculated as $c = \omega/k$ , the slope of the linear dispersion. To find out $c$, acoustic branches in the $\Gamma$-$W$ and $\Gamma$-$L$ directions are considered.   Values of $c$  of three acoustic branches in two directions are averaged, which we denote by $c_{avg}$  and used to estimate $\kappa_{ph}$ instead of $c$ in Eq. 3. Table III shows the values of $c$  calculated for three acoustic branches in two directions and  $c_{avg}$ values for Fe$_{2}$ScX, Fe$_{2}$VAl and Fe$_{2}$TiSn  compounds. The branch notations followed are shown in the inset of Fig. 1 (b). 
 
 The Table III shows that the $c$ of phonons of three acoustic branches in $\Gamma$-$L$ directions are reducing from Fe$_{2}$ScP to Fe$_{2}$ScSb. In $\Gamma$-$W$ direction, similar trend in $c$ is noted for branch S2 and S3 (which are degenerate), except for S1 in which opposite trend is observed. In $\Gamma$-$L$ direction, branches S1 and S2 are degenerate and have value of $c$, 102.4, 86.5 and 78.5 $THz\cdot Bohr$ for Fe$_{2}$ScX, respectively. The $c_{avg}$ for Fe$_{2}$ScX are 112.2, 99.2, and 91.0 $THz\cdot Bohr$, respectively. The observed trend is due to reduction in phonon frequencies on going from first to last compound. In both directions  considered, $c$ values of phonons in Fe$_{2}$VAl is greater than that in Fe$_{2}$TiSn for all branches. The calculated $c_{avg}$ are 116.6 and 107.1 $THz\cdot Bohr$ for Fe$_{2}$VAl and Fe$_{2}$TiSn, respectively. The phonon $c_{avg}$ of Fe$_{2}$ScX compounds are lesser compared to Fe$_{2}$VAl and Fe$_{2}$TiSn except, for Fe$_{2}$ScP with $c_{avg}$ of 112.2 $THz\cdot Bohr$.

 Once, the quantities $c_{v}$ and $c$ are known, $\tau_{ph}$ needs to be found out which is the computationally more difficult part. At this step, to estimate the value of $\tau_{ph}$, we considered Fe$_{2}$VAl and later the $\tau_{ph}$ is benchmarked using Fe$_{2}$TiSn compound. For the estimation, $c_{v}$ values are taken from our previous work\cite{paper2} and $c_{avg}$ (and $c$) used are calculated and tabulated in Table III.  Using the experimental $\kappa$ of Fe$_{2}$VAl\cite{lue1} at 300 K of $\sim$25 $Wm^{-1}K^{-1}$,  the $\tau_{ph}$ estimated,  is $\sim$5x10$^{-13}s$. The calculated value of $\kappa_{ph}$ for Fe$_{2}$TiSn using the estimated $\tau_{ph}$ at the same temperature is $\sim$14.8 $Wm^{-1}K^{-1}$. This value is close to the experimental $\kappa$ value of Leu \textit{et al.}\cite{lue2}($\sim$7 $Wm^{-1}K^{-1}$) at 300 K. This suggests that the $\tau_{ph}$ estimated by this method can predict reasonably accurate values of $\kappa_{ph}$  for other three compounds using such a simple approach. It is important to note here that, Fe$_{2}$VAl, Fe$_{2}$TiSn and Fe$_{2}$ScX compounds have same crystal structure and belong to the same class of compounds with similar phonon dispersions which led us to select the former two compounds in the estimation. With the calculated $\tau_{ph}$ of 5x10$^{-13}s$ and using $c_{avg}$ and $c_{v}$ values, the estimated $\kappa_{ph}$ at 300 K for three compounds are 18.2, 13.6 and 10.3  $Wm^{-1}K^{-1}$, respectively. The $\kappa_{ph}$ is reducing from Fe$_{2}$ScP to Fe$_{2}$ScSb, which may be mainly due to the larger sized and heavier mass elements in the latter compounds.

\begin{table*}
\caption{Phase velocity ($c$) of three acoustic branches (S1-S3) in $\Gamma$-$W$ and $\Gamma$-$L$ directions and average of phase velocities, $c_{avg}$, for Fe$_{2}$ScX (X=P, As, Sb), Fe$_{2}$VAl and   Fe$_{2}$TiSn compounds in $THz\cdot Bohr$}

\begin{tabular}{@{\extracolsep{\fill}} c c c c c c c c c c c c c c c c }
 \hline\hline

& & \multicolumn{2}{c}{Fe$_{2}$ScP} &  & \multicolumn{2}{c}{Fe$_{2}$ScAs} &  & \multicolumn{2}{c}{Fe$_{2}$ScSb} & & \multicolumn{2}{c}{Fe$_{2}$VAl} & & \multicolumn{2}{c}{Fe$_{2}$TiSn} \\
 \cline{3-4} \cline{6-7} \cline{9-10} \cline{12-13} \cline{15-16}
     &  &$\Gamma$-$W$ &$\Gamma$-$L$ & &$\Gamma$-$W$ &$\Gamma$-$L$ & &$\Gamma$-$W$ &$\Gamma$-$L$ & &$\Gamma$-$W$ &$\Gamma$-$L$ & &$\Gamma$-$W$ &$\Gamma$-$L$ \\ 
\hline     
S1 &   &69.9  &102.4 & &71.0  &86.5  & &73.4  &78.5  & &96.7   &101.4 & &80.1   &93.1  \\ 
S2 &   &95.5  &102.4 & &84.4  &86.5  & &78.1  &78.5  & &101.4  &101.4 & &90.4   &93.1  \\ 
S3 &   &161.9 &141.0 & &138.7 &128.3 & &120.6 &117.1 & &151.1  &147.8 & &146.2  &139.9 \\ 
\hline
$c_{avg}$ &  & &112.2 & & &99.2 & & &91.0 & & &116.6  &   & &107.1 \\
\hline
\end{tabular}
\end{table*}

The assessment of TE applicability of Fe$_{2}$ScX Heusler compounds is done by calculating temperature dependent $ZT$ in 300-1200 K  range. For the calculation of $ZT$ from Eq. 1, the estimated $\kappa_{ph}$ at 300 K of the pure compounds are used for all temperatures, which is generally expected to reduce at higher temperatures. Constant electron relaxation time $\tau$ of 1x10$^{-14}s$ is used\cite{ashcroft} in the estimation of $S^{2}\sigma$ and $\kappa_{e}$ . The temperature dependent $ZT$ values of Fe$_{2}$ScX compounds are calculated for the highest PF yielding optimal electron (n-type) doping  of $\sim$1x10$^{21}$ cm$^{-3}$  and hole (p-type) doping of $\sim$5x10$^{21}$ cm$^{-3}$,  out of the doping range considered in this study. The obtained $ZT$ for doped Fe$_{2}$ScX compounds are presented in Fig. 8. The $ZT$ values suggest that Fe$_{2}$ScX compounds could be promising materials for TE applications with $ZT$ showing the increasing nature with temperature. 

 From the figure we can see that for electron doping shown, Fe$_{2}$ScSb has the highest $ZT$ while, for hole doping  Fe$_{2}$ScP shows the highest $ZT$ among the family. The  highest $ZT$ is obtained from  Fe$_{2}$ScSb, which  in 300-500 K is  0.11-0.25, in the mid-temperature range of 500-900 K is 0.25-0.45 and in high temperature range 900-1200 K is 0.45-0.52. The $ZT$ values obtained from the hole doping of $\sim$5x10$^{21}$ cm$^{-3}$ for these compounds are lower compared to electron doping. The n-type Fe$_{2}$ScP has $ZT$ in the range 0.07 to 0.43, while p-type Fe$_{2}$ScP has 0.03-0.34 which is the highest among p-type Fe$_{2}$ScX compounds.  The doped Fe$_{2}$ScAs is showing intermediate $ZT$ among the optimal doped Fe$_{2}$ScX compounds. The $ZT$ range for n-type and p-type Fe$_{2}$ScAs are  0.09-0.48 and 0.03-0.28 in 300-1200 K, respectively. For last compound in the family, n-type Fe$_{2}$ScSb, the highest $ZT$ is observed in the range of 0.11-0.52. In case of it's p-type counterpart of $\sim$5x10$^{21}$ cm$^{-3}$ doped, $ZT$ at 300 K is 0.03 and it is reaching 0.26 at 1200 K. Higher  $ZT$ values are observed in the high temperature region for both type of doping, which hints at the possible applicability in high temperature power generation. The 900-1200 K range $ZT$ for $\sim$1x10$^{21}$ cm$^{-3}$ doping of Fe$_{2}$ScX compounds are, 0.34-0.43, 0.40-0.48 and 0.45-0.52, respectively. For the  hole doping of $\sim$5x10$^{21}$ cm$^{-3}$, the $ZT$ values in the same temperature range are 0.25-0.34, 0.20-0.28 and 0.18-0.26, respectively.

The higher $ZT$ observed at temperatures $>$900 K, indicates that for the  high temperature TE power generation, doped Fe$_{2}$ScX full-Heuslers could be one of the suitable candidates. Currently, doped Si-Ge alloys are one of TE materials in use for power generation in high temperature region with $ZT$ ranges in 673-1273 K of $\sim$0.5-0.8 and $\sim$0.4-0.6 for n-type and p-type Si-Ge, respectively\cite{snydercomplex}. The simultaneous increase in thermopower and electrical conductivity with moderate decrease in thermal conductivity  was observed in the work of Makongo \textit{et al.}\cite{nanostruct}  leading to $\sim$2.5 times $ZT$ enhancement for the bulk nanostructured (Zr,Hf)NiSn half-Heusler alloys. In the TE research, with the considerable interest in enhancement of $ZT$ by complex nanostructuring\cite{snydercomplex,nano2}, we anticipate that Fe$_{2}$ScX family of full-Heusler could be worth cosidering next to Si-Ge alloys. Also, the $ZT$ plot, hints that, with both electron and hole doped compounds showing good $ZT$ number, they can be used as n-type and p-type thermoelements in constructing TEGs. Therefore, an attempt by experimental community to synthesize and study the TE properties are highly desired.

\begin{figure}
\vspace{1.5cm}
\includegraphics[width=8cm, height=5cm]{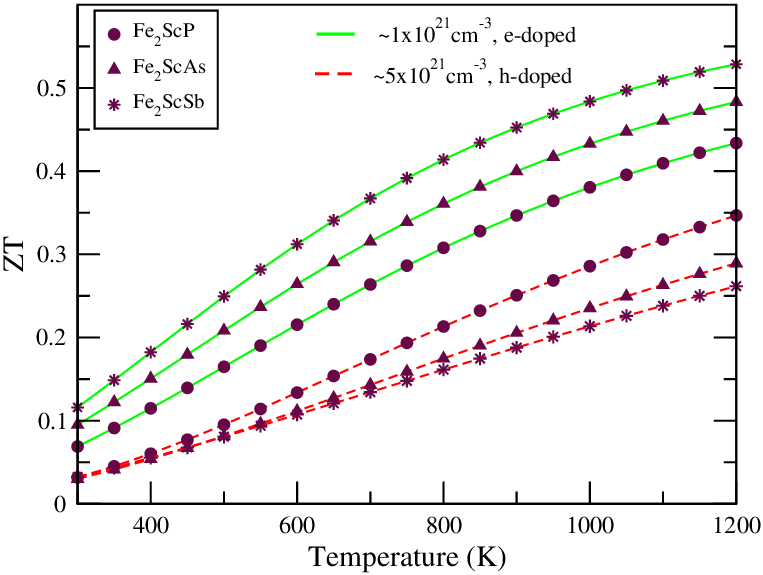} 
\caption{Temperature dependent $ZT$ plot of Fe$_{2}$ScX (X=P, As, Sb) compounds for electron doping of $\sim$1x10$^{21}$ cm$^{-3}$ shown in solid (green) lines and hole doping of $\sim$5x10$^{21}$ cm$^{-3}$ in dashed (red) lines. }

\end{figure}

\section{Conclusions} 
In the present work, electronic structure and dependent thermoelectric coefficients of Fe$_{2}$ScX (X=P, As, Sb) compounds are studied using two functionals \textit{viz.} SCAN and mBJ. The differences in the features of the electronic bands obtained from mBJ and SCAN are brought out by comparison. The effective mass values are calculated at the bands extrema of SCAN. The indirect band gaps for Fe$_{2}$ScP, Fe$_{2}$ScAs and Fe$_{2}$ScSb, obtained from mBJ are 0.81, 0.69 and 0.60 $eV$, respectively. $S$, ${\sigma}/{\tau}$, PF and ${\kappa_{e}}/{\tau}$ are calculated for various values of electron and hole doping by introducing band gaps from mBJ into band structures of SCAN under semiclassical transport equations. The PF obtained at the temperature of 800 K for electron doping of $\sim$1x10$^{21}$ cm$^{-3}$ are  $\sim$121, $\sim$125 and $\sim$131 $10^{14}\mu WK^{-2}cm^{-1}s^{-1}$  for Fe$_{2}$ScX where  X= P, As, Sb, respectively. While, at the same temperature,  highest PF values is obtained for hole doping of $\sim$5x10$^{21}$ cm$^{-3}$ are $\sim$118, $\sim$83 and $\sim$63 $10^{14}\mu WK^{-2}cm^{-1}s^{-1}$  for Fe$_{2}$ScX, respectively. The phonon dispersion and density of states are calculated for the three compounds which supported the structural stability. The calculated lattice contribution to $c_{v}$ shows at any temperature $c_{v}$ are in order Fe$_{2}$ScSb $>$ Fe$_{2}$ScAs $>$ Fe$_{2}$ScP. The maximum phonon energy (or frequency) for Fe$_{2}$ScX are $\sim$55, $\sim$48 and $\sim$43 meV, respectively. The Debye temperatures are calculated for three compounds which are $\sim$637 K, $\sim$556 K and $\sim$498 K, respectively. Based on the calculation of slopes of acoustic branches in the linear region and estimated $\tau_{ph}$,  $\kappa_{ph}$ of Fe$_{2}$ScX compounds at 300 K are estimated. The $\kappa_{ph}$ for three compounds are 18.2, 13.6 and 10.3 $Wm^{-1}K^{-1}$, respectively. In the temperature range 900-1200 K, the $ZT$ values are 0.25-0.34, for the  hole doping of $\sim$5x10$^{21}$ cm$^{-3}$ are  0.20-0.28 and 0.18-0.26, respectively. In the same temperature range, $ZT$ for $\sim$1x10$^{21}$ cm$^{-3}$ doping are 0.34-0.43, 0.40-0.48 and 0.45-0.52, respectively. Our work suggests that Fe$_{2}$ScX compounds could be worth considering for high temperature TE applications with both n-type and p-type compounds showing good $ZT$ values.

\section{Acknowledgements}
The authors thank Science and Engineering Research Board (SERB), Department of Science and Technology, Government of India for funding this work. This work is funded under the SERB project sanction order No. EMR/2016/001511.

\section{References}
\bibliography{ref}
\bibliographystyle{apsrev4-1}


\end{document}